# Post-orientation free single molecule imaging by XFELs

Zoltán Jurek and Gyula Faigel

*Wigner Research Centre for Physics, Institute for Solid State Physics and Optics, POB 49, 1525 Budapest, Hungary*

**Abstract** – Single molecule imaging is one of the main target areas of X-ray free electron lasers. It relies on the possibility of orienting the large number of low counting statistics 2D diffraction patterns taken at random orientations of identical replicas of the sample. This is a difficult process and the low statistics limits the usability of orientation methods and ultimately it could prevent single molecule imaging. We suggest a new approach, which avoids the orientation process from the diffraction patterns. In our schema one measures the orientation of the sample together with the diffraction pattern by detecting some fragments of the Coulomb explosion. We show by molecular dynamics simulations that from the angular distribution of the fragments one can obtain the orientation of the samples.

**Introduction.** – Recently, the first hard XFEL source LCLS started operation, and others are under construction. The unique characteristics of these sources open new possibilities for structural studies. One of these is single molecule imaging [1]. The basic idea is that one collects 2D diffraction patterns from a large number of identical replicas of the sample and compiles these into a 3D diffraction pattern and reconstructs the atomic structure from this. There are many difficulties in carrying out this type of measurements. Among those we mention two: the radiation damage of individual samples and the orientation of 2D diffraction patterns, because these are relevant to our present study. Calculations show that the samples turn into plasma in a short time. Structural data have to be collected before this happens, before the original atomic structure deteriorates. Many models have been worked out to describe the atomic motions in the sample [2,3,4,]. All models agree that pulse length of 10 fs or shorter is necessary to avoid appreciable deterioration of the sample and to reach atomic resolution. Few years ago, so short pulses were a dream. However, with the startup of LCLS the first results suggest that the pulse length can be shortened by decreasing the charge in the electron bunches. This low charge operation mode is already included in the baseline design of the XFEL-s under construction, and the minimum pulse length is set around 2 fs for these devises. With these short pulses, radiation damage can be circumvented. The other large obstacle is the unknown orientation of the consecutively introduced samples into the XFEL beam. To compile the 3D diffraction pattern we have to know the orientation of the individual samples arriving at the probe beam. In practice the orientation cannot be measured before the particles arrive at the beam. Therefore one has to figure it out for the millions of 2D diffraction patterns after the data collection. There are various approaches [5, 6, 7, 8,9,10] to do this. However, these have limitations: like minimum number of photons in a pixel (statistical requirements), scalability (calculation complexity increases very strongly with the number of pixels and with the number of 2D patterns), further it is not proved that for large realistic systems these iterative methods converge to the proper fix-point at all. The most serious of the above problems is the statistical requirement. There is a minimum number of photons in a single picture, which allow these methods to work. Shortening the pulse length may solve the problem of radiation damage but at the same time, it makes the determination of the orientation more difficult, because short pulses are at a price. The price is the low bunch charge, which leads to a low number of photons in a pulse. An other consequence of the statistical requirements is a limitation on sample size. This comes from the fact that the total elastic scattering cross section for small samples (below a few thousand atoms) are so small that even from a well focused intense pulse the number of elastically scattered photons are very low. So the individual patterns may not have good enough statistics to carry enough information for orientation. Moreover, even if a method seemed to work, it would be important to check its result by an independent way. Therefore large efforts are concentrated to find some independent source of additional information on the orientation of the samples. One possibility is to orient the samples in advance, before they arrive at the probe pulse. Such methods are under development. However, it is clear that these will also be restricted to samples with high anisotropy, and the orientation of the sample will not be very precise. Here we suggest another approach, which relies on the analyses of the particle distribution of the exploding sample. Experimental setups capable of measuring the spatial and velocity distribution of fragments are already available. In this paper we show that for samples containing only a few heavy atoms, the orientation of the sample can be determined by measuring the angular distribution of these atoms after the Coulomb explosion of the sample. Although this method cannot be used for all samples in general, we suggest an extension, which significantly enlarge the type of samples measurable this way.

[a]E-mail: faigel.gyula@wigner.mta.hu

**Modelling and analyses**

This For analyzing the particle distributions and for showing that our suggestion works we use our molecular dynamics type modeling tool developed for the description of the Coulomb explosion of particles in an intense hard x-ray beam [3, 11].
In the model the motion of all particles are followed by solving the non-relativistic equations of motion. Coulomb forces are explicitly included, and the various quantum-processes are taken into account through their cross sections as stochastic processes. In the homogeneous sample limit our model gives similar results as simple continuum models [2], but with more details [3]. In the non-homogeneous case, which is relevant for biological systems, and the continuum approach cannot be properly used, our model predicts significant deviations from the behavior of homogeneous samples [12]. In the present study we use the unique feature of our model: the possibility to calculate the behavior of non-homogeneous systems. Further we extend the calculations for much longer time scales then previous works, which modeled the sample dynamics during the time of the x-ray pulse only. The reason of this extension is that we intend to model the angular distribution of fragments as they arrive at far away detectors, i.e. long time after the x-ray pulse.
The model calculations were done for typical hard XFEL parameters: 12 keV energy, $10^{13}$ photons/pulse fluence, 10 fs flat top shape pulse and 100 nm diameter focal spot. For a clear illustration of our idea, first we use a simple model system; a homogeneous spherical carbon sample containing 7200 C atoms and 3 Fe atoms (Fe1=outer, Fe2=middle, and Fe3=inner, see fig.1.a) at different distances from the center of the sample. The density of the molecule is 1.35 g/cm$^3$ typical for biological systems. The structure of the sample is shown in fig. 1.

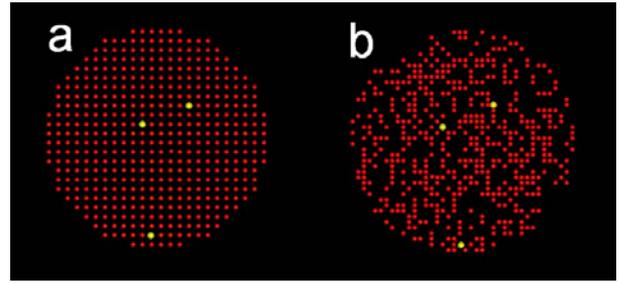

Fig. 1: Cross section of the homogeneous (a) and inhomogeneous model samples. Red dots are Carbon atoms, while yellow dots are Fe atoms. Fe3 is closest to the centre, Fe2 is halfway F1 is close to the perimeter of the molecule.

We chose this configuration in order to illustrate how heavy atoms behave at different places of the samples. We also present model calculations for an inhomogeneous sample, which mimic biological systems well [12]. All types of calculations were repeated at least 10 times with different random numbers, to see how stable the results are. The time span of the calculations was about 300 fs. We have checked that after this time the particles moved on straight lines, so their positions on the detector can be precisely determined. This is illustrated in fig.2. where the three components of the velocity unit vectors of Fe1 and Fe2 sites are depicted.

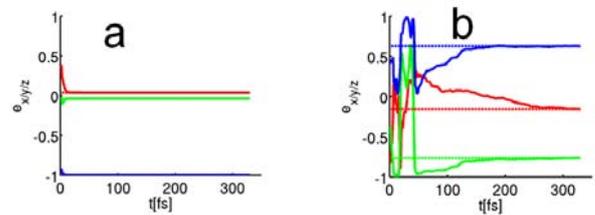

Fig. 2: Typical time evolution of the components of the velocity unit vectors for Fe1 (a) and Fe3 sites (b). After 300 fs there is no appreciable variation of any components.

Following the time evolution of the Coulomb explosion of the homogeneous sample we find that there is no appreciable atomic motion in the first 5 fs, at 10 fs atomic displacements in the outer shell of the sample show up, and at 50 fs the original structure is fully destroyed and the size of the atomic cloud is much larger then the original size of the sample, and this continues as the time goes on. This behavior agrees with our earlier findings [3,11,12]. Now we concentrate on the motion of Fe atoms, since we would like to use them to fix the orientation of the molecule. For this reason we plotted the outgoing directions of the three different Fe sites in fig 3. A clear trend can be observed: the atoms closer to the perimeter of the sample have a narrower outgoing angular distribution. Those atoms, which are only one-two atomic layer below the surface, leave the sample within a 2 degrees wide cone (fig3.a). Atoms coming almost from the center of the sample leave the system in a much wider (about 12 degrees) angular range (fig3.c). Atoms at halfway to the surface of the molecule leave the sample between the above two values in about a 5 degrees cone.

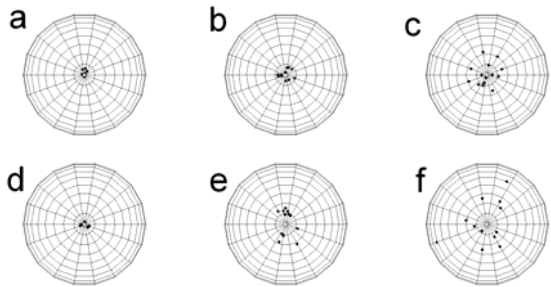

Fig. 3: Angular distribution of Fe atoms for the homogeneous (upper panel) and inhomogeneous (lower panel) samples. The three different iron sites (Fe1, Fe2 and Fe3) are shown on (a,b,c) and (d,e,f) respectively.

The same calculations were done for an inhomogeneous system (fig1.b). This sample was built in the following way: first C atoms were placed on a dense regular grid a=1.5 Å. This would lead to 6.25 g/cm$^3$ density. To reach the 1.35 g/cm$^3$ target density typical for biological specimens, randomly chosen atoms were removed, resulting in the structure shown in fig1.b. The outgoing directions of Fe atoms are plotted in fig.3.d,e,f. The trend is the same as for homogeneous samples, and it is graphically shown in fig.4. for both samples.

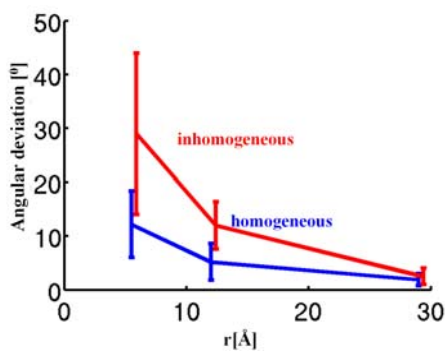

Fig. 4: Width of the heavy atom angular distribution as a function of distance from the centre of the sample, red inhomogeneous and blue homogeneous sample.

In the inhomogeneous case, the distributions widened in general. The atoms closest to the center have a very wide (30 degrees) distribution, but the atoms at the perimeter have a 2.5 degrees wide distribution, which leaves us with a quite definite outgoing direction. If we use this direction for orienting the sample we can fix one rotation axis by having a single site with heavy element. This in itself reduces the search space from 3D to 1 D, which is a great help for data evaluation. However, we may have an even better situation with 2 or 3 heavy atoms in the sample at well defined sites. In the case of two different heavy atoms the orientation of the molecule can be determined almost uniquely with an ambiguity of a mirror symmetry. Having 3 different heavy atoms in general positions (not too close to each other and close to the perimeter of the sample) the orientation of the molecule is fully determined. So there is no need to calculate the orientation posteriori from the 2D diffraction patterns. In the case of having 2 or 3 heavy atoms of the same type, we can also gain similar orientation information with some restriction. If the angular distances between pairs are not equal the identification of the orientation can be done.

At last we would like to point out that orientation information from heavy atom identification and angular distribution can be gained not only for special molecules containing heavy atoms but for a much wider class of samples. In biology there are well developed methods to add tags for definite sites of various species (viruses, proteins etc). These tags can contain heavy elements, and therefore they could be used for orientation purposes. The above approach would allow the precise determination of the 3D orientation of molecules without the use of the diffraction pattern. Further, using this scheme would lift many requirements on pulse parameters such as pulse length, and fluence, and it would allow the measurement of smaller samples. So single molecule imaging could be done for a much wider variety of molecules and for less stringent beam conditions then thought before.

**Summary** – We have shown that in single molecule imaging experiments, orientation information can be obtained from the measurement of heavy atom angular distribution. For molecules having a single heavy atom close to the surface of the molecule one rotation axis is fixed, resulting in a reduction of the orientation search space from 3D to 1D. Having two different heavy atoms in the sample the orientation is determined with the exception of mirror symmetry, while the orientation is uniquely determined for molecules with 3 different heavy atoms. Similarly, attaching three marker molecules containing three different heavy atoms to the outer surface of the sample we can determine the 3D orientation of the molecule solely from the measurement of their angular distribution. In the last two cases the x-ray diffraction patterns do not have to be used at all for gaining orientation information. This means that the requirements on pulse length, on the number of photons in a pulse, and even on the minimum molecular size are relaxed. So using our molecular dynamics modeling tool we have shown that measuring the fragment distribution in parallel with the x-ray pattern in a single molecule imaging experiment, we can significantly expand the possibilities of these experiments and we might be able to carry out these studies on a much wider variety of molecules and with less stringent beam conditions then earlier expected.

\*\*\*

Acknowledgments

The authors thank János Hajdú for initiating this study, and Abrahám Szőke and Miklós Tegze for the illuminating discussions. This work has been funded by grants OTKA, 105691 and K81348.